# Monitoring Software Reliability using Statistical Process Control: An Ordered Statistics Approach

Bandla Srinivasa Rao
Associate Professor.,
Dept. of Computer Science
VRS & YRN College

Dr. R Satya Prasad
Associate Professor,
Dept. of Computer Science & Eng.,
Acharya Nagarjuna University

Dr. R.R. L Kantham
Professor,
Dept. of Statistics
Acharyan Nagarjuna University

## ABSTRACT
The nature and complexity of software have changed significantly in the last few decades. With the easy availability of computing power, deeper and broader applications are made. It has been extremely necessary to produce good quality software with high precession of reliability right in the first place. Olden day's software errors and bugs were fixed at a later stage in the software development. Today to produce high quality reliable software and to keep a specific time schedule is a big challenge. To cope up the challenge many concepts, methodology and practices of software engineering have been evolved for developing reliable software. Better methods of controlling the process of software production are underway. One of such methods to assess the software reliability is using control charts. In this paper we proposed an NHPP based control mechanism by using order statistics with cumulative quantity between observations of failure data using mean value function of exponential distribution.

## 1. INTRODUCTION
As computer applications became more diverse and spread through almost every area of everyday life, reliability became a very important characteristic for software, since it is a matter of economy. To produce a software having reliability, it is necessary to measure and control its reliability. To do this, a number of models have been developed; new models try to make better predictions. Software reliability represents a user oriented view of software quality. It relates directly to operation rather than design of the program, and hence it is dynamic. For this reason software reliability is interested in failures occurrence and not faults in a program

### 1.1. Software reliability Modeling
The probability that a given program will work as intended by the user, i.e., without failures in a specified environment and for a specified duration can be termed as software reliability [1][2]. The aim of software engineer is to increase this probability and make it one if possible. To do this one must measure the reliability of the software. A commonly used approach for measuring software reliability is by using an analytical model whose parameters are generally estimated from available data on software failures. Reliability quantities have been defined with respect to time, although it is possible to define them with respect to other variables. We have taken inter failures time data of Musa(1975) which are random values. In reliability study there are two characteristics of a random process: 1) the probability distribution of the random variables, i.e., Poisson and 2) the variation of the process with time. A random process whose probability distribution varies with time is called non homogeneous. For the random process for time variation we can define two functions, the mean value function m(t), as the average cumulative failures associated with each time point and the failure intensity function $\lambda$, as the rate of change of mean value function. When there are changes in the software i.e. software corrections occur it is called non homogeneous process.

Let M(t) be the random process representing the number of failures experienced by time t, then the mean value function is defined by $\mu(t) = E[m(t)]$. i.e. the expected number of failures at time t. the failure intensity function of the $m(t)$ process is the instantaneous rate of change of the expected number of failures with respect to time or $\lambda(t) = \frac{d\mu(t)}{dt}$. [3]

## 2. ORDERED STATISTICS
Let X denote a continuous random variable with Probability Density Function (PDF) f(x) and Cumulative Distribution Function (CDF) F(x), and let $(X_1, X_2, …, X_n)$ denote a random sample of size n drawn on X. The original sample observations may be unordered with respect to magnitude. A transformation is required to produce a corresponding ordered sample. Let $(X_1, X_2, …, X_n)$ denote the ordered random sample such that $X_1 < X_2 < … < X_n$; then $(X_1, X_2, …, X_n)$ are collectively known as the order statistics derived from the parent X. The various distributional characteristics can be known from Balakrishnan and Cohen [4]. The inter-failure time data represent the time lapse between every two consecutive failures. On the other hand if a reasonable waiting time for failures is not a serious problem, we can group the inter-failure time data into non 29overlapping successive sub groups of size 4 or 5 and add the failure times within each sub group. For instance if a data of 100 inter-failure times are available we can group them into 20 disjoint subgroups of size 5. The sum total in each subgroup would denote the time lapse between every 5$^{th}$ order statistics in a sample of size 5. In general for inter-failure data of size 'n', if r (any natural no) less than 'n' and preferably a factor n, we can conveniently divide the data into 'k' disjoint subgroups (k=n/r) and the cumulative total in each subgroup indicate the time between every r$^{th}$ failure. The probability distribution of such a time lapse would be that of the r$^{th}$ ordered statistics in a subgroup of size r, which would be equal to r$^{th}$ power of the distribution function of the original variable m (t). The whole process involves the mathematical model of the mean value function and knowledge about its parameters. If the parameters are known they can be taken as they are for the further analysis, if the parameters are not know they have to be estimated using a sample data by any admissible, efficient method of distribution. This is essential because the control limits depend on mean value function, which intern depends on the parameters. If software failures are quite frequent keeping track of inter-failure is tedious. If failures are more frequent order statistics are preferable.[5]

## 2.1. Model Description

Considering failure detection as a non homogenous Poisson process with an exponentially decaying rate function, the expected number of failures observed by time t is given by $m(t) = a(1 - e^{-bt})$ and the failure rate by $\lambda(t) = m'(t)$. To calculate the parameter values and control limits using Order Statistics approach, we considered exponential distribution [8]. The mean value function of exponential distribution is

$$m(t) = a(1 - e^{-bt})$$

In order to group the inter-failure time data into non overlapping successive sub groups of size r the mean value function can be written as

$$m(t) = a(1 - e^{-bt})^r$$

$$m(s_k) = [a(1 - e^{-bs_k})]^r$$

$$m'(s_k) = a^r r (1 - e^{-bs_k})^{r-1} b \, e^{-bs_k} \qquad 2.1.1$$

The likelihood function L can be written as

$$L = e^{-m(s_n)} \prod_{k=1}^{n} m'(s_k) \qquad 2.2.2$$

Substituting eq-2.1.1 in eq-2.2.2 we can write

$$L = e^{-m(s_n)} \prod_{k=1}^{n} a^r r (1 - e^{-bs_k})^{r-1} b \, e^{-bs_k}$$

$$\log L = -m(s_n) + \sum_{k=1}^{n} [\log a^r + \log b + \log r + \log e^{-bs_k} + \log(1 - e^{-bs_k})^{r-1}] \qquad 2.2.3$$

$$m(s_n) = [a(1 - e^{-bs_n})]^r \qquad 2.2.4$$

Substitute equation 2.2.4 in 2.2.3 we get

$$\log L = -[a(1 - e^{-bs_n})]^r + \sum_{k=1}^{n} [\log a^r + \log b + \log r + \log e^{-bs_k} + \log(1 - e^{-bs_k})^{r-1}]$$
2.2.5

$$\frac{\partial \log L}{\partial a} = 0,$$
$$a^r = \frac{n}{(1 - e^{-bs_n})^r} \qquad 2.2.6$$

$$\frac{\partial \log L}{\partial b} = 0,$$
$$a^r r \, s_n \, e^{-bs_n} (1 - e^{-bs_n})^{r-1} + \frac{n}{b} + 0 \quad - \sum_{k=1}^{n} s_k + (r-1) \sum_{k=1}^{n} \frac{s_k e^{-bs_k}}{(1 - e^{-bs_k})} \qquad 2.2.7$$

Substitute equation 2.2.6 in 2.2.7 we obtain the following equation

$$g(b) = \frac{nr s_n e^{-bs_n}}{(1 - e^{-bs_n})} + \frac{n}{b} - \sum_{k=1}^{n} s_k + (r-1) \sum_{k=1}^{n} \frac{s_k e^{-bs_k}}{(1 - e^{-bs_k})} \qquad 2.2.8$$

Derivate with respect to b of equation 2.2.8 we obtain

$$g'(b) = \frac{-nr s_n^2 e^{-bs_n}}{(1 - e^{-bs_n})^2} - \frac{n}{b^2} - (r-1) \sum_{k=1}^{n} \frac{(s_k^2 e^{-bs_k})}{(1 - e^{-bs_k})^2} \qquad 2.2.9$$

## 2.2. Parameter estimation and Control limits

Parameter estimation is a statistical method trying to estimate parameters based on inter failures time data which is based on ordered statistics. For the given observations using equations 2.2.8 and 2.2.9 the parameters 'a' and 'b' are computed by using the popular Newton Rapson method A program written in C was used for this purpose. [3]

Based on the time between failures data given in Table-1, we compute the software failure process through mean value control chart. We use cumulative time between failures data for software reliability monitoring through SPC. The parameters obtained from Goel-Okumoto model applied on the given time domain data are as follows:

**Table 1: Parameter estimates and their control limits of 4 and 5 order Statistics**

| | Order | $\hat{a}$ | $\hat{b}$ |
|---|---|---|---|
| Data Set of Table 2 | 4$^{th}$ | 2.415117 | 0.000099 |
| | 5$^{th}$ | 1.933309 | 0.000114 |

'$\hat{a}$' and '$\hat{b}$' are ordered statistics of parameters and the values can be computed using analytical method for the given time between failures data shown in Table 1. Using values of 'a' and 'b' we can compute $m(t)$. Now equate the *pdf* of m*(t) to* 0.00135, 0.99865, and 0.5 and the respective control limits are given by

$$T_U = (1 - e^{-bt}) = 0.99865$$

International Journal of Computer Applications (0975 – 8887)
Volume 32– No.7, October 2011

$$T_C = (1 - e^{-bt}) = 0.5$$

$$T_L = (1 - e^{-bt}) = 0.00135$$

These limits are convert at $m(t_u)$, $m(t_c)$ and $m(t_L)$ are given by

$$m(t_u) = 33.3512569382986,$$

$$m(t_c) = 16.6981710073481,$$

$$m(t_L) = 0.04508506100108$$

They are used to find whether the software process is in control or not by placing the points in Mean value chart shown in figure-1.and figure-2. A point below the control limit $m(t_L)$ indicates an alarming signal. A point above the control limit $m(t_u)$ indicates better quality. If the points are falling within the control limits it indicates the software process is in stable. [6]

## STATISTICAL PROCESS CONTROL

Statistical process control is the application of statistical methods to provide the information necessary to continuously control or improve processes throughout the entire lifecycle of a product [7]. SPC techniques help to locate trends, cycles, and irregularities within the development process and provide clues about how well the process meets specifications or requirements. They are tools for measuring and understanding process variation and distinguishing between random inherent variations and significant deviations so that correct decisions can be made about whether to make changes to the process or product. One of such primary statistical technique used to assess process variation is the control chart. [8]

### 2.3. Control Chart

The control chart displays sequential process measurements relative to the overall process average and control limits. The upper and lower control limits establish the boundaries of normal variation for the process being measured. Variation within control limits is attributable to random or chance causes, while variation beyond control limits indicates a process change due to causes other than chance, a condition that may require investigation. [7] The upper control limit (UCL) and lower control limit (LCL) give the boundaries within which observed fluctuations are typical and acceptable There are many different types of control charts, pn, p, c, etc., [8], [9],[10]

### 2.4. Developing Control Chart

Given the n inter-failure data the values of m(t) at $T_c$, $T_u$, $T_L$ and at the given n inter-failure times are calculated. Then successive differences of m(t)'s are taken, which leads to n-1 values. The graph with the said inter-failure times 1 to n-1 on X-axis, the n-1 values of successive differences m(t)'s on Y-axis, and the 3 control lines parallel to X-axis at $m(T_L)$, $m(T_U)$, $m(T_C)$ respectively constitutes mean value chart to assess the software failure phenomena on the basis of the given inter-failures time data.

### 2.5. Illustration

The procedure of a mean value chart for failure software process will be illustrated with an example here. Table 1 show the time between failures of software product reported by Musa (1975) [11].

**Table 2: Software failure data reported by Musa (1975) [11]**

| Fault | Time | Fault | Time | Fault | Time | Fault | Time | Fault | Time | Fault | Time | Fault | Time | Fault | Time |
|---|---|---|---|---|---|---|---|---|---|---|---|---|---|---|---|
| 1 | 3 | 18 | 120 | 35 | 227 | 52 | 21 | 69 | 529 | 86 | 860 | 103 | 108 | 120 | 22 |
| 2 | 30 | 19 | 26 | 36 | 65 | 53 | 233 | 70 | 379 | 87 | 983 | 104 | 0 | 121 | 75 |
| 3 | 113 | 20 | 114 | 37 | 176 | 54 | 134 | 71 | 44 | 88 | 707 | 105 | 3110 | 122 | 482 |
| 4 | 81 | 21 | 325 | 38 | 58 | 55 | 357 | 72 | 129 | 89 | 33 | 106 | 1247 | 123 | 5509 |
| 5 | 115 | 22 | 55 | 39 | 457 | 56 | 193 | 73 | 810 | 90 | 868 | 107 | 943 | 124 | 100 |
| 6 | 9 | 23 | 242 | 40 | 300 | 57 | 236 | 74 | 290 | 91 | 724 | 108 | 700 | 125 | 10 |
| 7 | 2 | 24 | 68 | 41 | 97 | 58 | 31 | 75 | 300 | 92 | 2323 | 109 | 875 | 126 | 1071 |
| 8 | 91 | 25 | 422 | 42 | 263 | 59 | 369 | 76 | 529 | 93 | 2930 | 110 | 245 | 127 | 371 |
| 9 | 112 | 26 | 180 | 43 | 452 | 60 | 748 | 77 | 281 | 94 | 1461 | 111 | 729 | 128 | 790 |
| 10 | 15 | 27 | 10 | 44 | 255 | 61 | 0 | 78 | 160 | 95 | 843 | 112 | 1897 | 129 | 6150 |
| 11 | 138 | 28 | 1146 | 45 | 197 | 62 | 232 | 79 | 828 | 96 | 12 | 113 | 447 | 130 | 3321 |
| 12 | 50 | 29 | 600 | 46 | 193 | 63 | 330 | 80 | 1011 | 97 | 261 | 114 | 386 | 131 | 1045 |
| 13 | 77 | 30 | 15 | 47 | 6 | 64 | 365 | 81 | 445 | 98 | 1800 | 115 | 446 | 132 | 648 |
| 14 | 24 | 31 | 36 | 48 | 79 | 65 | 1222 | 82 | 296 | 99 | 865 | 116 | 122 | 133 | 5485 |
| 15 | 108 | 32 | 4 | 49 | 816 | 66 | 543 | 83 | 1755 | 100 | 1435 | 117 | 990 | 134 | 1160 |
| 16 | 88 | 33 | 0 | 50 | 1351 | 67 | 10 | 84 | 1064 | 101 | 30 | 118 | 948 | 135 | 1864 |
| 17 | 670 | 34 | 8 | 51 | 148 | 68 | 16 | 85 | 1783 | 102 | 143 | 119 | 1082 | 136 | 4116 |

**Table 3:4<sup>th</sup> order cumulative faults and their m(t) successive difference.**

| Fault | 4-order Cumulative Faults | m(t) | Successive Difference's of m(t) | Fault | 4-order Cumulative Faults | m(t) | Successive Difference's of m(t) |
|---|---|---|---|---|---|---|---|
| 1 | 227 | 0.053669607 | 0.050189929 | 18 | 16358 | 1.936901718 | 0.083134596 |
| 2 | 444 | 0.103859536 | 0.070964302 | 19 | 18287 | 2.020036314 | 0.079829413 |
| 3 | 759 | 0.174823838 | 0.064912354 | 20 | 20567 | 2.099865727 | 0.09363885 |
| 4 | 1056 | 0.239736192 | 0.191343658 | 21 | 24127 | 2.193504577 | 0.077302531 |
| 5 | 1986 | 0.431079851 | 0.131004192 | 22 | 28460 | 2.270807108 | 0.046687097 |
| 6 | 2676 | 0.562084043 | 0.296000509 | 23 | 32408 | 2.317494206 | 0.039546532 |
| 7 | 4434 | 0.858084551 | 0.097761832 | 24 | 37654 | 2.357040738 | 0.020362766 |
| 8 | 5089 | 0.955846384 | 0.042703058 | 25 | 42015 | 2.377403504 | 0.001034693 |
| 9 | 5389 | 0.998549441 | 0.132378119 | 26 | 42296 | 2.378438197 | 0.016427908 |
| 1 | 6380 | 1.13092756 | 0.12873383 | 27 | 48296 | 2.394866105 | 0.006274847 |
| 11 | 7447 | 1.259661391 | 0.053077534 | 28 | 52042 | 2.401140952 | 0.001810038 |
| 12 | 7922 | 1.312738924 | 0.22760806 | 29 | 53443 | 2.40295099 | 0.003163639 |
| 13 | 10258 | 1.540346985 | 0.075916166 | 30 | 56485 | 2.406114629 | 0.004113047 |
| 14 | 11175 | 1.616263151 | 0.102288223 | 31 | 62651 | 2.410227676 | 0.000973225 |
| 15 | 12559 | 1.718551373 | 0.061080293 | 32 | 64893 | 2.411200901 | 0.00261935 |
| 16 | 13486 | 1.779631666 | 0.103253053 | 33 | 76057 | 2.413820251 | 0.000925177 |
| 17 | 15277 | 1.882884719 | 0.054016999 | 34 | 88682 | 2.414745429 | |

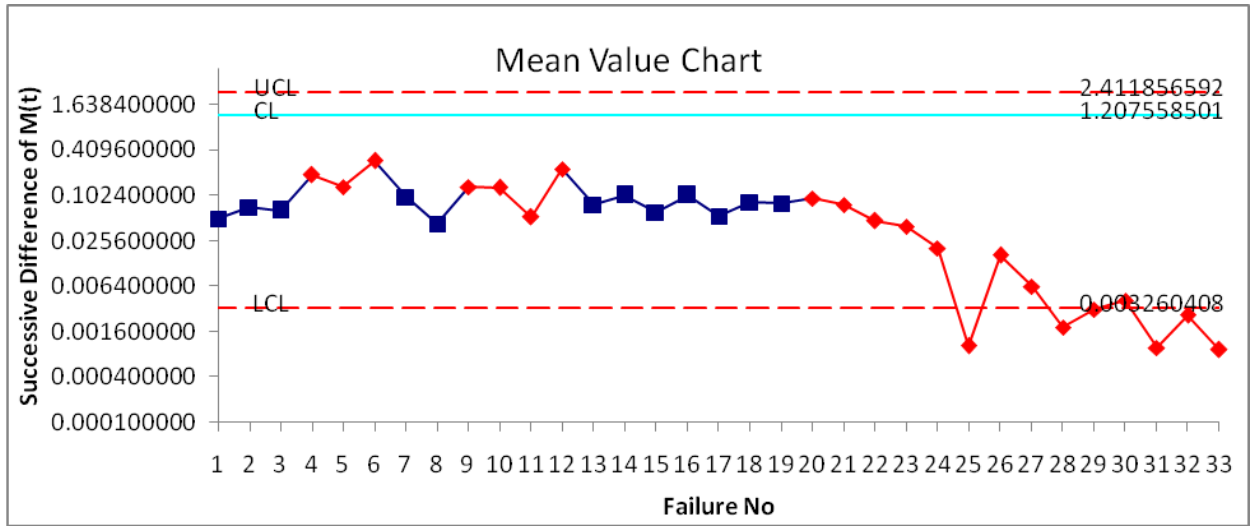

Fig-1: Mean Value Chart of 4<sup>th</sup> order data set

Table: 4: 5<sup>th</sup> order cumulative faults and their m(t) successive difference

| Fault | 5-order Cumulative | m(t) | Successive Difference's Of m(t)'s | Fault | 5-order Cumulative | m(t) | Successive Difference's Of m(t)'s |
|---|---|---|---|---|---|---|---|
| 1 | 342 | 0.073925386 | 0.04791294 | 15 | 17758 | 1.62782472 | 0.050149302 |
| 2 | 571 | 0.121838326 | 0.080156008 | 16 | 20567 | 1.677974022 | 0.069965628 |
| 3 | 968 | 0.201994334 | 0.189702018 | 17 | 25910 | 1.74793965 | 0.084558265 |
| 4 | 1986 | 0.391696352 | 0.183547444 | 18 | 29361 | 1.832497915 | 0.032788872 |
| 5 | 3098 | 0.575243796 | 0.270820129 | 19 | 37642 | 1.865286786 | 0.041557817 |
| 6 | 5049 | 0.846063925 | 0.033556388 | 20 | 42015 | 1.906844603 | 0.010389187 |
| 7 | 5324 | 0.879620314 | 0.11950945 | 21 | 45406 | 1.91723379 | 0.005154027 |
| 8 | 6380 | 0.999129764 | 0.125362531 | 22 | 49416 | 1.922387817 | 0.004007073 |
| 9 | 7644 | 1.124492295 | 0.196749357 | 23 | 53321 | 1.92639489 | 0.002484134 |
| 10 | 10089 | 1.321241652 | 0.059242997 | 24 | 56485 | 1.928879023 | 0.001341437 |
| 11 | 10982 | 1.380484649 | 0.090964117 | 25 | 62661 | 1.930220461 | 0.001561037 |
| 12 | 12559 | 1.471448766 | 0.100355072 | 26 | 74364 | 1.931781497 | 0.001125183 |
| 13 | 14708 | 1.571803838 | 0.056020882 | 27 | 84566 | 1.93290668 | 0.00027658 |
| 14 | 16185 | 0.073925386 | 0.04791294 | | | | |

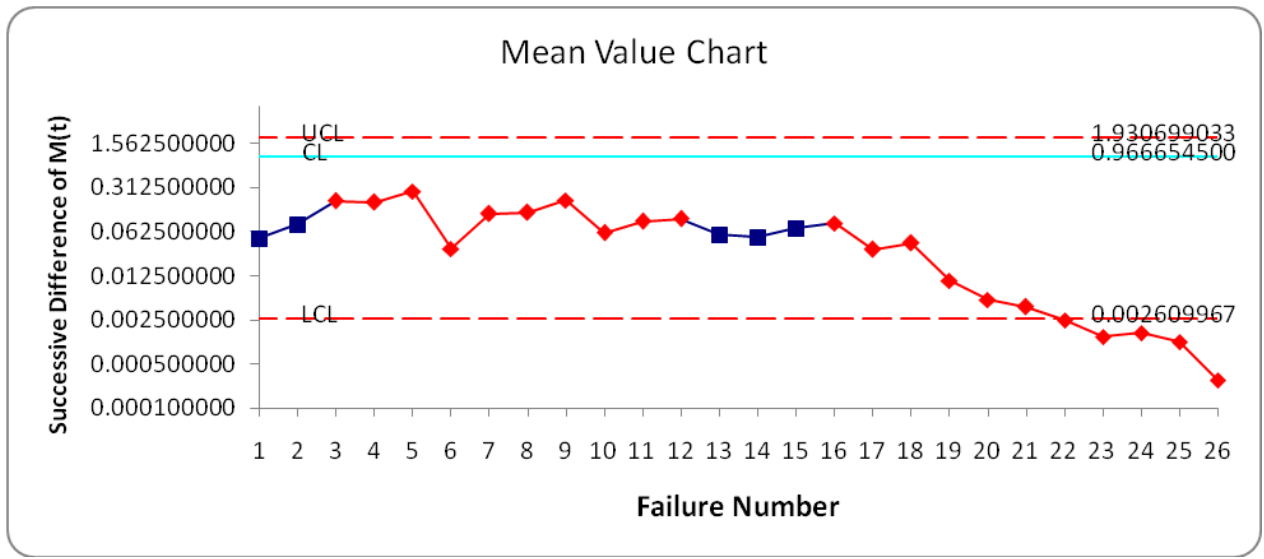

**Fig 2:** Mean Value Chart of 5[th] order data set

## 3. CONCLUSION

The Mean value charts of Fig 1 and 2 have shown out of control signals i.e. below LCL. By observing Mean value charts, we identified that failures situation is detected at an early stages. The early detection of software failure will improve the software reliability. When the control signals are below LCL, it is likely that there are assignable causes leading to significant process deterioration and it should be investigated. Hence, we conclude that our control mechanism proposed in this chapter giving a positive recommendation for its use to estimate whether the process is in control or out of control.

## 4. REFERENCES

[1] Musa, J.D, Software Reliability Engineering McGraw-Hill, 1998

[2] Musa, J.D., Iannino, A., Okumoto, k., 1987. "Software Reliability: Measurement Prediction Application". McGraw-Hill, New York.

[3] Pham. H., 2003. "Handbook of Reliability Engineering", Springer

[4] Balakrishnan.N., Clifford Cohen; Order Statistics and Inference; Academic Press inc.;1991.

[5] K.Ramchand H Rao, Dr. R.Satya Prasad, Dr. R.R.L.Kantham, Assessing Software Reliability Using SPC : An Order Statistics Approach, International Journal of Computer Science, Engineering and Applications (IJCSEA) Vol.1, No.4, August 2011

[6] MacGregor, J.F., Kourti, T., 1995. "Statistical process control of multivariate processes". Control Engineering Practice Volume 3, Issue 3, March 1995, Pages 403-414

[7] The Organizational Process Management Cycle Programmed Workbook, Interaction Research Institute, Inc., Fairfax, Virginia.

[8] Carleton, A.D. and Florac, A.W. 1999. Statistically controlling the Software process. The 99 SEI Software Engineering Symposimn, Software Engineering Institute, Carnegie Mellon University

[9] Smith, Gary, Statistical Reasoning, Allyn and Bacon, Boston, MA, 1991.

[10] Caprio, William H., "The Tools for Quality," Total Quality Management Conference, Ft. Belvoir, Virginia, July 13-15, 1992.

[11] Juran, J. M. (ed.), Juran's Quality Control Handbook, 4th ed., McGraw-Hill, Inc., New York, 1988.

[12] Hong Pharm; System Reliability; Springer;2005;Page No.281